\documentstyle[12pt]{article}
\input{psfig}
\begin{document}

\title{A link of information entropy and kinetic energy for quantum 
many-body systems.}

\author{S.E. Massen and C.P. Panos\\
Department of Theoretical Physics\\
Aristotle University of Thessaloniki,
54006 Thessaloniki, Greece }

\maketitle

\begin{abstract}

A direct connection of information entropy $S$ and kinetic energy $T$ 
is obtained for nuclei and atomic clusters, which establishes $T$ as a
measure of the information in a distribution. It is conjectured
that this is a universal property for fermionic many-body systems. We
also check rigorous inequalities previously found to hold between S
and T for atoms and verify that they hold for nuclei and atomic clusters
as well. These inequalities give a relationship of Shannon's information
entropy in position-space with an experimental quantity i.e. the rms
radius of nuclei and clusters.
\end{abstract}

\vspace*{1cm}

Information entropy is an important entity employed for the study of quantum-
mechanical systems \cite{Bialy75,Gadre84,Gadre85,Gadre87,Ohya93,%
Nagy96,Majer96,Panos97,Lalazi98,Massen98,Panos00}.
In \cite{Bialy75} an entropic uncertainty relation (EUR)
was discovered, which for a three-dimensional system has the form:

\begin{equation}
S = S_{r}+S_{k}{\ge} 3(1+\ln{\pi}) \cong 6.434,\quad (\hbar =1)
\label{SrSk-1}
\end{equation}
where $S_r$ is the information entropy in position-space:
\begin{equation}
S_{r}=-{\int}{\rho}({\bf r}) \ln {\rho}({\bf r})d {\bf r}
\label{Sr}
\end{equation}
and the information entropy in momentum-space $S_k$ is:
\begin{equation}
S_{k}=-{\int}n({\bf k})\ln n({\bf k})d{\bf k}
\label{Sk}
\end{equation}
$\rho({\bf r})$ and $n({\bf k})$ are the density distributions in 
position- and momentum- space respectively, normalized to one. However, 
for a normalization to  the number of particles $N$, the following EUR 
holds \cite{Gadre87}:
\begin{equation}
S = S_{r}+S_{k}{\ge} 3 N (1+\ln{\pi}) -2N \ln N
\label{SrSk-2}
\end{equation}
The above inequalities are an expression of quantum-mechanical uncertainty
in conjugate spaces, since an increase of $S_k$ is accompanied by a
decrease of $S_r$ and vice versa, which indicates that a diffuse $n(k)$
is associated with a localised $\rho(r)$ and vice versa. This is expected
by Heisenberg's uncertainty relations, although EUR are stronger than them.
$S_r$ and $S_k$ depend on the unit of length in measuring $\rho({\bf r})$ 
and $n({\bf k})$ respectively. However, the
important quantity is the entropy sum $S_r + S_k$ (net information content
of the system) which is invariant to uniform scaling of coordinates.

In \cite{Massen98} we proposed a universal property for $S$ for the total 
density distributions of nucleons in nuclei, electrons in atoms and valence 
electrons in atomic clusters.
This property has the form:
\begin{equation}
S=a  + b  \ln N
\label{S-ab-1}
\end{equation}
where the parameters $a$, $b$  depend on the system under consideration.

Relation (\ref{S-ab-1}) holds for density distributions normalized to unity.
However, for the aim of the present work, we transform 
(\ref{S-ab-1}) for
normalization to the number of particles $N$
\begin{equation}
S=a N + b N \ln N 
\label{S-ab-2}
\end{equation}

The above transformation can be easily done using the 
relation \cite{Massen98}:
\begin{equation}
S[norm=1]=\frac{S[norm=N]}{N} + 2 \ln N  
\label{S-norm}
\end{equation}
Thus relation (\ref{S-ab-1}) is transformed to 
$  S=a N + (b-2) N \ln N $
which putting $b-2 \rightarrow b$ gives relation (\ref{S-ab-2}).
 
Equation (\ref{S-norm}) can be proved as follows: Let $\rho(r)$ be the
density distribution normalized to one and  $\rho_N(r)= N \rho(r)$
the corresponding one normalized to $N$. Then
\begin{eqnarray}
S_r[norm=N] &=& - \int \rho_N(r) \ln \rho_N(r) d {\bf r}
=- \int N \rho(r) (\ln N  + \ln \rho(r)) d {\bf r}
\nonumber\\
\nonumber\\
&=&-N \ln N + N S_r[norm=1]
\label{Sr-norm}
\end{eqnarray}
Similarly
\begin{equation}
S_k[norm=N] = -N \ln N + N S_k[norm=1]
\label{Sk-norm}
\end{equation}

Combining Eqs. (\ref{Sr-norm}) and  (\ref{Sk-norm}) and taking into account
that $S=S_r +S_k$ we obtain Eq. (\ref{S-norm}).

In ref. \cite{Lalazi98} we employed another definition of entropy according 
to phase-space considerations. Thus we derived an information-theoretic 
criterion of the quality of a nuclear density distribution i.e. the 
larger $S$, the better the quality of $\rho(r)$.

The question, however, arises naturally how to connect $S$ with fundamental
quantities. Recently \cite{Panos00}, we considered the single-particle states
of a nucleon in nuclei, a $\Lambda$ in hypernuclei and an electron in atomic
clusters. We connected $S$ with the energy $E$ of single-particle
states through the relation
\begin{equation}
S= k \ln (\mu E + \nu)
\label{S-E}
\end{equation}
where $k$, $\mu$ and $\nu$ depend on the system. It is remarkable that 
the same relation holds for various systems.

For the total densities, there is already in atomic physics a connection of 
$S_r$ and $S_k$ with the total kinetic energy $T$ through rigorous 
inequalities derived using the EUR \cite{Gadre87}.
\begin{eqnarray}
\label{Sr-ineq}
S_r(min) & \leq & S_r \leq S_r(max) \\
\label{Sk-ineq}
S_k(min) & \leq & S_k \leq S_k(max) \\
\label{S-ineq}
S(min) & \leq & S \leq S(max)
\end{eqnarray}
where
\begin{eqnarray}
S_r(min)&=& \frac{3}{2}N(1+\ln \pi) + \frac{1}{2}N \ln N  
-\frac{3}{2} N \ln \left( \frac{4}{3}T \right) \nonumber\\
S_r(max)&=& \frac{3}{2}N(1+\ln \pi) + \frac{3}{2}N\ln 
\left( \frac{2}{3}<r^2> \right)
-\frac{5}{2} N\ln N  
\label{Sr-min}
\end{eqnarray}
\begin{eqnarray}
S_k(min)&=& \frac{3}{2}N(1+\ln \pi) + \frac{1}{2}N\ln N  
-\frac{3}{2}N\ln \left( \frac{2}{3}<r^2> \right) \nonumber\\
S_k(max)&=& \frac{3}{2}N(1+\ln \pi) - \frac{5}{2}N \ln N
+ \frac{3}{2}N\ln \left( \frac{4}{3}T \right)
\label{Sk-min}
\end{eqnarray}
\begin{eqnarray}
S(min)&=& 3 N(1+\ln \pi) - 2N\ln N   \nonumber\\
S(max)&=& 3 N(1+\ln \pi) +  \frac{3}{2}N
\ln \left( \frac{4}{9}<r^2><p^2> \right) - 5 N\ln N  
\label{S-min}
\end{eqnarray}

In the present work we verify numerically that the same inequalities hold 
for nuclear distributions and valence electron distributions in atomic 
clusters (see Tables \ref{table-1} and \ref{table-2} for some cases). 
For nuclei we employed 
Hartree-Fock wave functions obtained with SKIII interaction \cite{Dover72}.
In this model protons and neutrons move in different potentials. We choose
to work with the total density of the nucleons. However, similar results 
can be obtained for protons or neutrons.
For atomic (neutral sodium) clusters we employed the Woods-Saxon potential 
parametrized in \cite{Ekardt84,Kotsos97}.

It is seen from Tables \ref{table-1} and \ref{table-2}, however, that for 
$N$ large the inequalities are loose (the same was observed for atoms
\cite{Gadre87}), they are tighter
for $N$ small and perhaps can be made tighter by using more moments.
It is remarkable that the right-hand side of inequality (\ref{Sr-ineq})
is nearly an equality (comparison of columns 3 and 4 in Tables
\ref{table-1} and \ref{table-2}) 
and gives a relationship of Shannon's information
entropy in position-space $S_r$ with an experimental quantity i.e.
the rms radius of nuclei and clusters.

In Ref. \cite{Tao-Li}, $S_r$ and $S_k$ were connected with the first moments
$<r>$ and $<p>$ through similar inequalities which for atomic systems are
sharper than inequalities (\ref{Sr-ineq}), (\ref{Sk-ineq}) and
(\ref{S-ineq}). However this is not the case for nuclei and atomic clusters.
We tested numerically the inequalities of Ref. \cite{Tao-Li} and we found that
in these systems they hold but are not tighter than (\ref{Sr-ineq}), 
(\ref{Sk-ineq}) and (\ref{S-ineq}).

In the present Letter we proceed a step forward and establish a new link
of $S$ and $T$ for the total density distributions of nuclei and atomic
clusters.
We fitted our numerical values for several nuclei and atomic clusters
to the form:
\begin{equation}
S=\alpha T +\beta T \ln (\gamma T)
\label{S-T}
\end{equation}
Thus we obtained the following values of the parameters,
\[ \alpha = 0.4461 \ {\rm MeV
}^{-1}, \quad
   \beta = -0.0586\ {\rm MeV}^{-1}, \quad
   \gamma = 1\ {\rm MeV}^{-1}  \]
for nuclei and
\[ \alpha = 3.4722 \ {\rm eV}^{-1}, \quad
   \beta = -0.5970\ {\rm eV}^{-1}, \quad
   \gamma = 1\ {\rm eV}^{-1}  \]
for atomic clusters. The dependence of $S$ on $T$ is shown in fig. 1.
It is seen that expression (\ref{S-T}) represents the data well.
If this relationship proves to hold for other systems as well (e.g. atoms)
we might conjecture that this is a universal property.

We can justify easily the choice of (\ref{S-T})
by the following simple argument:
Our numerical calculations showed the relation $T\simeq cN$ ($c=$constant). 
Thus, our expression (\ref{S-ab-2}) $S=S(N)$ can be directly transformed 
to a similar relation $S=S(T)$. 

$T$ as function of $N$ according to our calculations is shown for nuclei
and atomic clusters in Fig. 2. Relation $T\simeq cN$ 
for the total kinetic energy or $T/N \simeq c$ shows that the
kinetic energy per particle for nuclei and atomic clusters is
approximately a constant. This holds for a quantum many body system 
of fermions moving in a mean field.

Concluding, in the present Letter we obtain a link between the
quantum-mechanical kinetic energy and the Shannon information entropy
for the total densities of nuclei and atomic clusters. We conjecture
that this is a universal property of fermionic many-body systems.
The idea that kinetic energy is related to the concept of information
should not strike one as radical. We quote \cite{Sears80}, where 
Fisher's information of a quantum many-body system is discussed and 
$T$ is expressed as a sum of two information functionals. It is stated in 
\cite{Sears80} that in broad terms one can always think of kinetic energy 
as a randomizing (entropic) force (as opposed to potential energy, which
exhibits constraining properties). Information theory simply provides a
mathematical link between the two concepts.

\newpage

\begin{table}
\caption{Values of $S_r$, $S_k$ and $S$ versus the number of particles
(nucleons) $N$ for nuclei using HF calculations with SKIII interaction.
Their lower and upper bounds are also shown.}
\label{table-1}
\begin{center}
\hspace*{-2cm}\begin{tabular}{ r r r r r r r r r r}
\hline
$N$ & $S_r(min)$& $S_r\,\,\,$& $S_r(max)$& $S_k(min)$& $S_k\,\,\,$
& $S_k(max)$& $S(min)$ &$S\,\,\,\,$& $S(max)$ \\
\hline
 4
 & -7.765  & 17.703 & 17.799 & -3.152  & -2.064 & 22.412
 & 14.646 & 15.639 & 24.453\\
 12 
 &-47.194  & 34.807 & 34.928 &-17.356  & -3.925 & 64.766
 & 17.573 & 30.882 & 75.808\\
 16 
 &-65.493  & 44.817 & 45.175 &-30.951  &-11.483 & 79.717
 & 14.224 & 33.333 &100.488\\
 24 
 & -112.571  & 63.898 & 64.802 &-62.928  &-21.857 &114.445
 &  1.874 & 42.041 &158.028\\
 28 
 & -136.528  & 72.429 & 73.770 &-80.217  &-28.940 &130.082
 & -6.446 & 43.489 &185.832\\
 32 
 & -158.467  & 82.916 & 84.787 & -100.700  &-39.195 &142.554
 &-15.913 & 43.722 &213.378\\
 40 
 & -204.612  &101.889 &104.244 & -141.987  &-59.821 &166.869
 &-37.743 & 42.068 &267.433\\
 50 
 & -272.574  &123.460 &127.454 & -196.946  &-80.417 &203.082
 &-69.493 & 43.043 &343.054\\
 70 
 & -403.488&170.777 &176.976 & -321.372& -137.777 &259.092
 & -144.396 & 33.000 &489.535\\
 90
 & -544.077&214.160 &223.862 & -454.751& -197.552 &313.188
 & -230.889 & 16.608 &640.155\\
 116
 & -731.653&273.758 &287.440 & -643.907& -283.525 &375.187
 & -356.467 & -9.767 &840.113\\
 208
 &-1442.694&477.190 &508.154 &-1390.258& -622.130 &560.590
 & -882.104 & -144.941 & 1570.384\\
\hline
\end{tabular}
\end{center}
\end{table}

\begin{table}
\caption{The same as in Table \ref{table-1} but for atomic clusters using 
a Woods-Saxon potential. Here $N$ is the number of valence electrons.}
\label{table-2}
\begin{center}
\hspace*{-2cm}\begin{tabular}{ r r r r r r r r r r}
\hline
$N$ & $S_r(min)$& $S_r\,\,\,$& $S_r(max)$& $S_k(min)$& $S_k\,\,\,$& $S_k(max)$& 
$S(min)$ &$S\,\,\,\,$& $S(max)$ \\
\hline
  4& 
    2.060& 17.774& 17.794&    -3.148&  0.396& 12.586&
   14.646& 18.170& 26.511\\
  8& 
   -1.277& 32.651& 32.883&  -14.681& -5.025& 19.479&
   18.203& 27.626& 52.941\\
 18& 
  -18.715& 71.144&  72.838& -61.076& -25.240& 30.477&
   11.762& 45.904& 126.514\\
  20&
  -22.565& 78.898&  80.264& -71.410& -30.390&  31.420&
    8.855& 48.508& 140.620\\
  58&
 -128.492& 223.477& 233.382& -331.211& -151.615&  30.664&
  -97.828&  71.862& 440.593\\
  68&
 -161.023& 262.366& 271.550& -407.878& -188.518&  24.695&
 -136.328&  73.849& 519.455\\
  70&
 -167.702& 270.139& 278.936& -423.332& -196.071&  23.306&
 -144.396& 74.069&  535.061\\
  92&
 -245.158& 353.016& 370.466& 
-610.530& -285.750& 5.094&
 -240.064&  67.266& 719.266\\
 126&
 -375.479& 483.600& 505.908& 
-913.943& -431.055& -32.557&
 -408.035&  52.546& 1003.518\\
 192&
 -650.053& 735.180& 773.811& 
-1557.324& -744.586& -133.461&
 -783.514&  -9.406& 1569.532\\
 198&
 -675.329& 757.645&  799.358& 
-1619.542& -776.252& -144.855&
 -820.184& -18.607& 1621.861\\
\hline
\end{tabular}
\end{center}
\end{table}

\begin{figure}
\label{fig-1}
\begin{center}
\begin{tabular}{cc}
{\psfig{figure=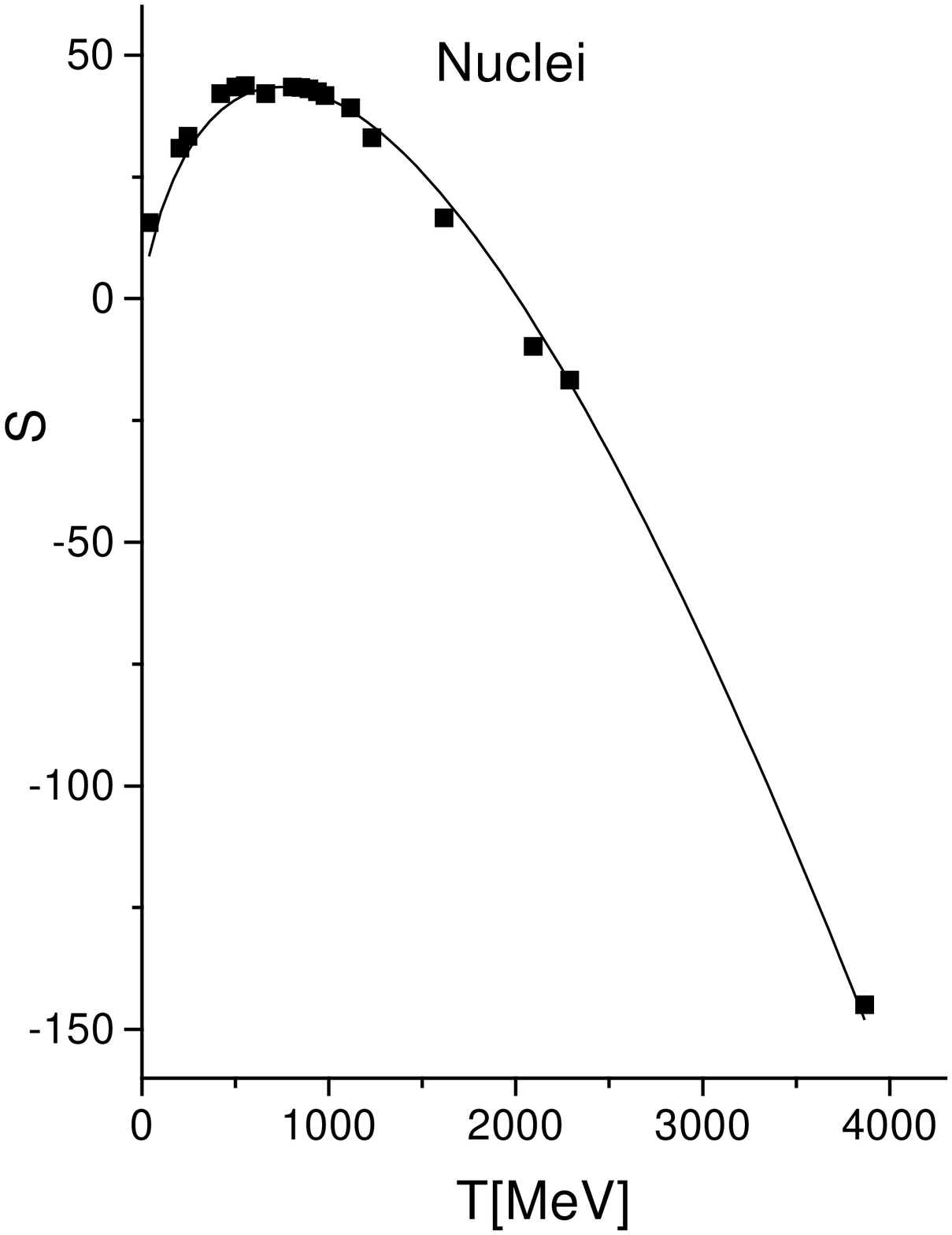,width=6.cm} }&
{\psfig{figure=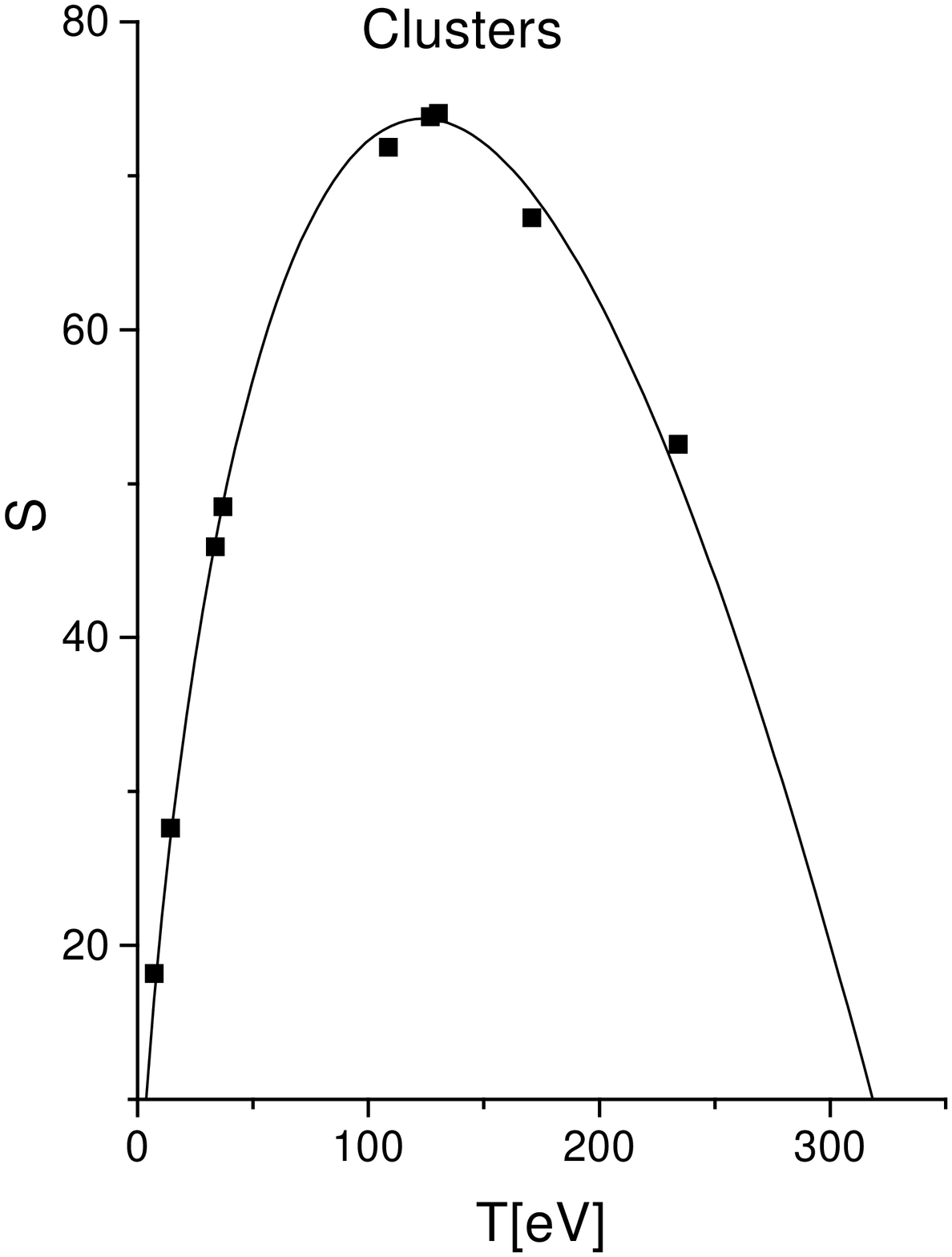,width=6.cm} }\\
\end{tabular}
\end{center}
\caption{Dependence of $S$ on $T$ for nuclei and atomic clusters.
The points correspond to our numerical calculations while the lines to our
fitted expressions (\ref{S-T}).}
\end{figure}
\begin{figure}
\label{fig-2}
\begin{center}
\begin{tabular}{cc}
{\psfig{figure=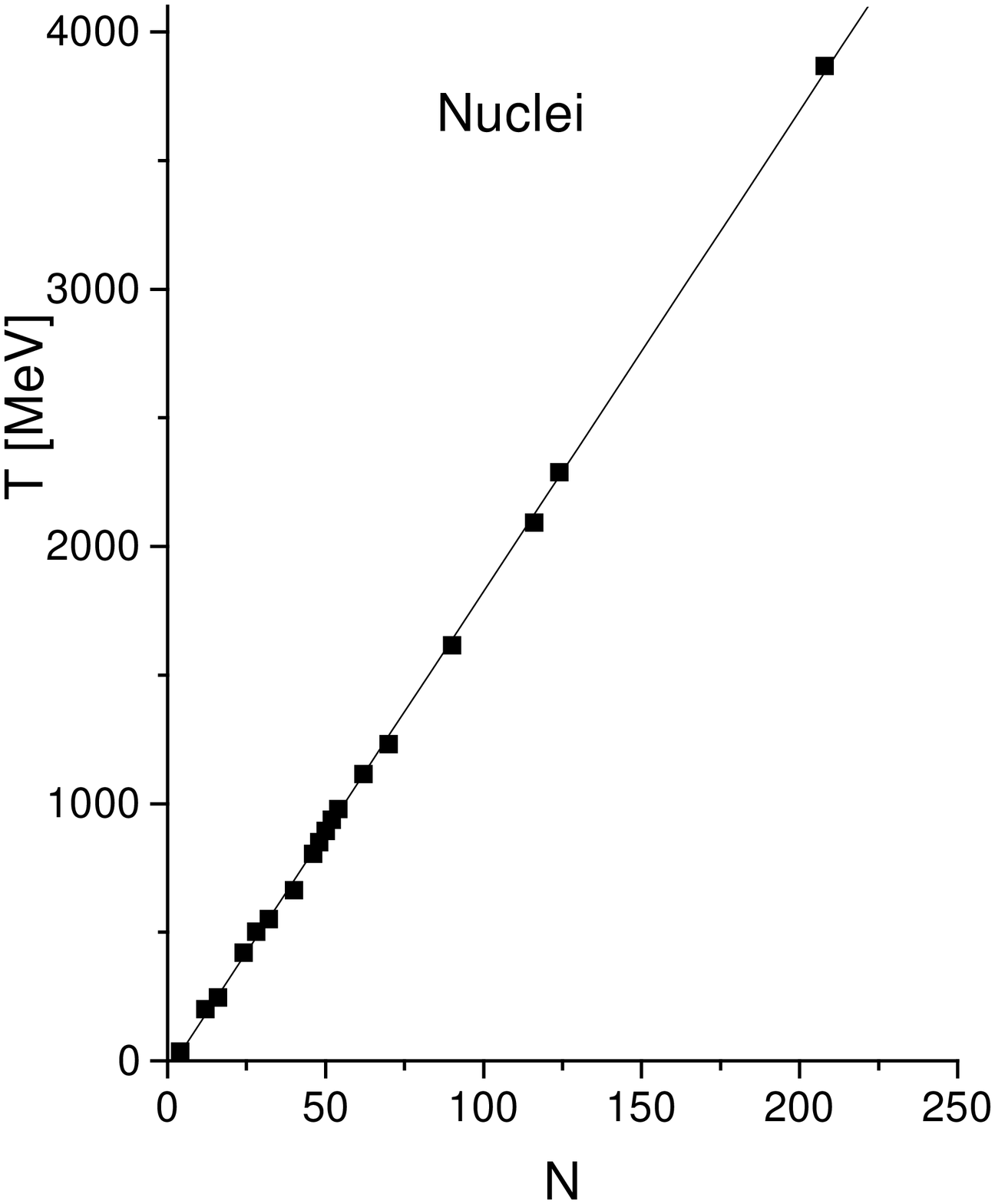,width=6.cm} }&
{\psfig{figure=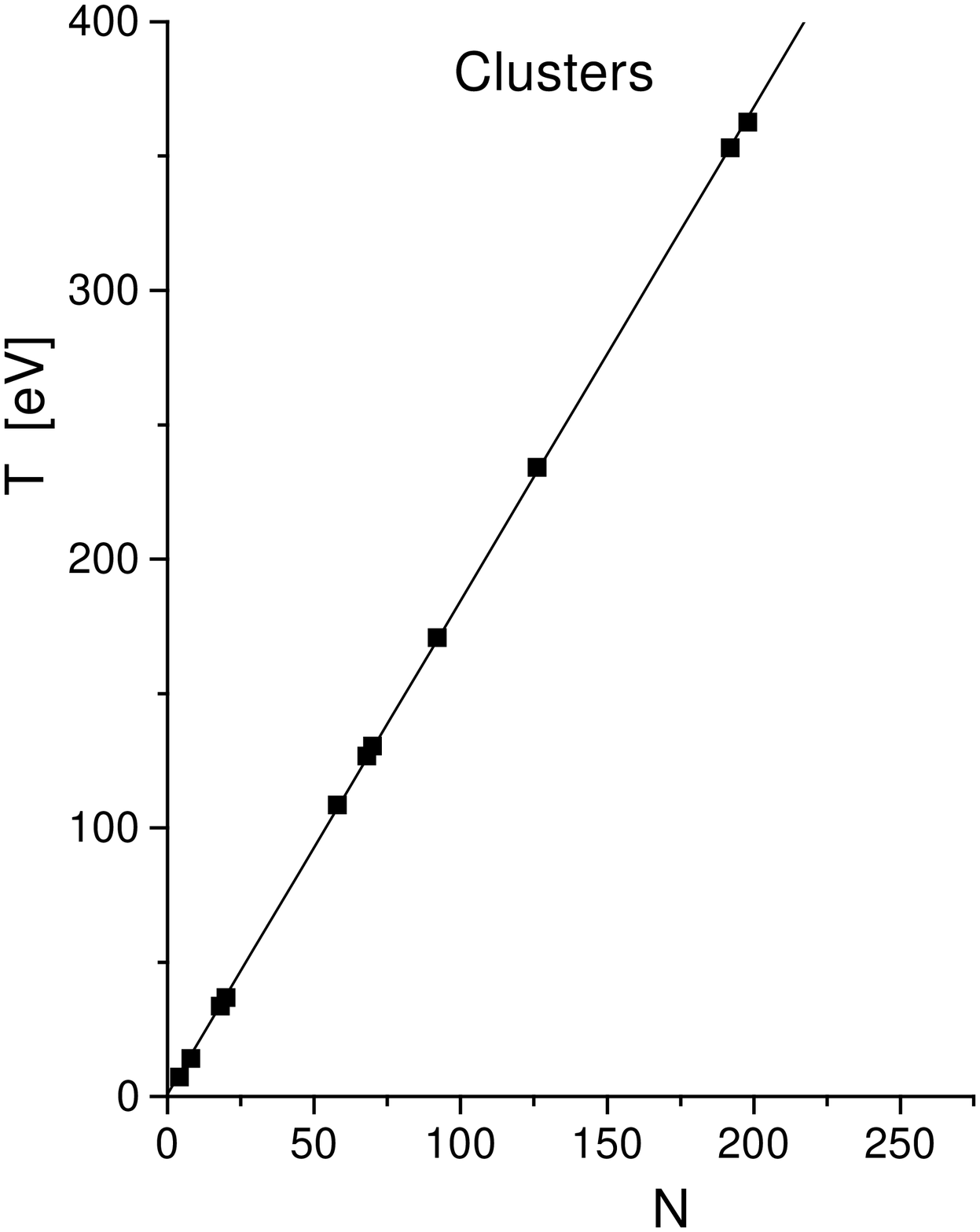,width=6.cm} }\\
\end{tabular}
\end{center}
\caption{Dependence of $T$ on the number of particles $N$ for nuclei and 
atomic clusters.
The points correspond to our numerical calculations while the lines to 
a least squares fit.}
\end{figure}

\end{document}